\documentclass[12pt]{article}
\usepackage[body={17cm, 23cm, centered}]{geometry}
\usepackage{amsmath,amssymb,amsthm,amscd,cite,comment,amsfonts,indentfirst,color,setspace,bbold,arydshln}
\usepackage{mathtext,marvosym,textcomp}
\usepackage[makeroom]{cancel}
\usepackage{mathtools}
\usepackage{cancel}

\usepackage[debug,pageanchor=false]{hyperref}
\hypersetup{colorlinks=true,linktocpage,breaklinks,
	urlcolor=blue,
	linkcolor=blue,
	citecolor=blue
}

\def\a{\alpha} 
\def\b{\beta} 
\def\g{\gamma} 
\def\d{\delta} 
\def\e{\epsilon}
\def\ve{\varepsilon} 
\def\z{\zeta} 
\def\h{\eta} 

\def\k{\kappa} 
\def\l{\lambda} 
\def\m{\mu}
\def\n{\nu} 
 
\def\r{\rho}
\def\q{\theta}

\def\f{\phi}
 
\def\w{\omega}

\def\L{\Lambda}

\def\W{\Omega}

\def\fr{\frac}  \def\dt{\partial}

\def\mc{\mathcal}

\def\mF{\mathcal{F}}

\def\mH{\mathcal{H}}
\def\mL{\mathcal{L}}

\def\mH{\mathcal{H}}

\def\mR{\mathcal{R}}

\def\tx{\tilde{x}}

\def\DD{{\mathcal{D}}}

\def\XX{\mathbb{X}}
\def\RR{\mathbb{R}}
\def\SS{\mathbb{S}}

\def\GL{\mathrm{GL}}
\def\SL{\mathrm{SL}}
\def\SO{\mathrm{SO}}

\selectfont

\begin{document}

\renewcommand{\refname}{\begin{center}References\end{center}}
	
\begin{titlepage}
		
	\vfill
	\begin{flushright}

	\end{flushright}
		
	\vfill
	
	\begin{center}
		\baselineskip=16pt
		{\Large \bf 
		Generalizing eleven-dimensional supergravity
		}
		\vskip 1cm
			Ilya Bakhmatov$^{a}$\footnote{\texttt{ibakhmatov@itmp.msu.ru}},
			Aybike Catal-Ozer$^{b}$\footnote{\texttt{ozerayb@itu.edu.tr}}, Nihat Sadik Deger$^{c}$\footnote{\texttt{sadik.deger@boun.edu.tr}}, Kirill Gubarev$^{d}$\footnote{\texttt{kirill.gubarev@phystech.edu}},\\ Edvard T. Musaev$^{d}$\footnote{\texttt{musaev.et@phystech.edu}}
		\vskip .3cm
		\begin{small}
			{\it 
                $^{a}$Institute of Theoretical and Mathematical Physics, Moscow State University, Russia \\
                $^{b}$Department of Mathematics, Istanbul Technical University, Istanbul, Turkey\\
                $^{c}$Department of Mathematics, Bogazici University,
                Bebek, 34342, Istanbul, Turkey\\
                $^{d}$Moscow Institute of Physics and Technology, Dolgoprudny, Russia
			}
		\end{small}
	\end{center}
		
	\vfill 
	\begin{center} 
		\textbf{Abstract}
	\end{center} 
	\begin{quote}
        We develop a procedure to reproduce the ten-dimensional generalized supergravity equations from T-duality covariant equations, that facilitates generalization to U-duality covariant formulations of eleven-dimensional supergravity.  The latter leads to a modification of the eleven-dimensional supergravity equations with terms that contain a rank-2 tensor field $J^{mn}$ which is the eleven-dimensional analog of the non-unimodularity Killing vector $I^m$ in ten dimensions.
	\end{quote} 
	\vfill
	\setcounter{footnote}{0}
\end{titlepage}
	
\clearpage
\setcounter{page}{2}
	
 
\section{Introduction}

An old and important problem in string/M-theory is description of consistent backgrounds for the fundamental string/membrane dynamics. A partial answer is given by the supergravity equations in 10 and 11 spacetime dimensions, which ensure kappa-symmetry of the Green--Schwarz (GS) superstring and of the membrane, respectively. At the quantum level, cancellation of the superstring Weyl anomaly is achieved by the supergravity as well. An intriguing observation made recently is that $\k$-symmetry of the GS superstring still holds if equations of motion for background fields are generalized to include a Killing vector $I^m$~\cite{Arutyunov:2015mqj, Wulff:2016tju}. This set of equations is usually referred to as generalized supergravity and has the following form:
\begin{equation}
    \begin{aligned}    
        R_{mn} - \fr14 H_{mpq}H_n{}^{pq}  +2 \nabla_{(m}Z_{n)} & =& T_{mn} ,\\
        -\fr12 \nabla^k H_{kmn} +Z^k H_{kmn} + 2\nabla_{[m}I_{n]} & =& K_{mn},\\ 
        R - \fr12 |H_3|^2 + 4 (\nabla^m Z_m - I^mI_m-Z^mZ_m) & =& 0,\\
        d *F_p - H_3 \wedge * F_{p+2} - \iota_I B_2\wedge *F_p -\iota_I *F_{p-2} &=& 0,
    \end{aligned}
    \label{eq:gensugra}
\end{equation}
where the expressions on the RHS read
\begin{equation}
    \begin{aligned}    
        T_{mn} & =&\fr14 e^{2\Phi}\sum_p\left[\fr1{p!}F_m{}^{k_1\dots k_{p}}F_{nk_1\dots k_{p}} - \fr12 g_{mn}|F_{p+1}|^2\right], \\
        K_{mn}   &=& \fr14 e^{2\Phi}\sum_p\fr1{p!}F_{k_1\dots k_p}F_{mn}{}^{k_1\dots k_p}.
    \end{aligned}
\end{equation}
Here $|\w_p|^2 = \frac{1}{p!} \w_{i_1\dots i_p}\w^{i_1\dots i_p}$ for a $p$-form $\w_p$, $Z_m=\dt_m\Phi + I^n B_{nm}$ and $I=I^m\dt_m$  is a Killing vector field that is also a symmetry of all fields in the theory, including the dilaton $\Phi$. Setting $I^m$ to zero one gets the usual equations of $d=10$ supergravity.

Although some subtleties arise at the quantum level related to the issue of locality of the generalized Fradkin--Tseytlin term~\cite{Fernandez-Melgarejo:2018wpg,Muck:2019pwj}, the classical $\kappa$--symmetry ensures one-loop scale invariance (UV finiteness) for a superstring propagating in generalized supergravity backgrounds~\cite{Arutyunov:2015mqj,Wulff:2016tju}. While they are lacking full conformal symmetry, their relation to consistent string theory backgrounds is known. Solutions to the generalized equations~\eqref{eq:gensugra} can be obtained from the standard supergravity solutions by T-duality transformations in the direction of an isometry that is broken by a linear dependence of the dilaton~\cite{Hoare:2015gda,Hoare:2015wia}. More generally, a sequence of (possibly non-abelian) T-dualities, or a Yang--Baxter (YB) deformation parametrized by a bivector $\b^{mn}= \fr12 r^{\a\b}k_\a{}^m k_\b{}^n$ is required. Here $k_\a{}^m$ are the Killing vectors of the \emph{initial} background and $r^{\a\b}=r^{[\a\b]}$ is a constant matrix (Greek indices label the isometry algebra elements, while small Latin indices refer to the spacetime coordinates). For the deformed background to be a solution of~\eqref{eq:gensugra} the $r$-matrix must satisfy the classical YB equation~\cite{Araujo:2017jkb,Bakhmatov:2017joy,Borsato:2018idb,Bakhmatov:2018bvp}
\begin{equation}
    \begin{aligned}
        r^{ \a[\g}r^{\d|\b|}f_{\a\b}{}^{\e]}&=0,
    \end{aligned}
    \label{cyb}
\end{equation}
where $f_{\a\b}{}^\g$ are the structure constants of the isometry algebra $[k_\a,k_\b]=f_{\a\b}{}^\g k_\g$. The vector $I^m$ is then simply 
\begin{equation} \label{I}
    I^m = r^{\a\b}f_{\a\b}{}^\g k_\g{}^m,
\end{equation}
and thus generalized supergravity solutions correspond to the non-unimodular YB deformations. The integrability preservation property of the YB deformation~\cite{Klimcik:2008eq} extends to generalized supergravity solutions, the renowned example being the ABF background~\cite{Arutyunov:2013ega,Arutyunov:2015qva}. 

Translating the above narrative into the 11-dimensional language encounters certain difficulties. In contrast to superstring,
there are no known integrability properties and no conformal versus scale invariance in the case of the supermembrane. Classical $\k$-symmetry of the membrane action leads to the torsion constraints that imply the standard 11d supergravity equations for the background fields~\cite{Bergshoeff:1987cm}. Both for the superstring and the supermembrane the $\k$-symmetry implies that the dimension $\fr12$ torsion component is expressed in terms of a spinor superfield $\chi_\a$. While in 10d, in addition to $\kappa$-symmetry, one may or may not require this to be a spinor derivative of the dilaton, $\chi_\a = \nabla_\a \Phi$, leading to the difference between the usual and generalized supergravities~\cite{Wulff:2016tju}, there is no such freedom in 11d as there is no dilaton. Instead, the scalar superfield $\Phi$ in this case can be gauged away by a super-Weyl transformation~\cite{Howe:1997he}.

In this letter we propose a possible way of circumventing these obstacles using the methods of exceptional field theories (ExFT, see \cite{Berman:2020tqn, Baguet:2015xha} for review). We show that the ExFT, yielding U-duality covariant formulations of 11-dimensional supergravity, provide a highly efficient tool for constructing a generalization of 11d supergravity equations, such that broken symmetry is the global GL(11) symmetry. We develop an algorithm that yields a set of generalized equations for the fields of 11d supergravity, which include an additional tensor $J^{mn}$, whose properties are very similar to those of the $I^m$ above \eqref{I}. Specifically, the 11d generalization of YB deformation, now parametrized by a tri-vector $\Omega^{mnk} = \frac{1}{3!} \rho^{\a\b\g}k_\a{}^mk_\b{}^nk_\g{}^k$ with some constant $\rho^{\a\b\g} = \r^{[\a\b\g]}$, is required to satisfy an analog of the unimodularity constraint in order to generate solutions of the standard supergravity equations~\cite{Gubarev:2020ydf}. This has the form
\begin{equation}\label{uni}
    J^{mn} = \rho^{\a\b\g} f_{\b\g}{}^{\d} k_{\a}{}^m k_{\d}{}^n = 0.
\end{equation}
Together with the so-called generalized YB equation \cite{Bakhmatov:2019dow} 
\begin{equation}
\label{eq:sl5YB}
    6 \rho^{\a\b[\g} \rho^{\d\e |\z|} f_{\a\z}{}^{\h]} + {\rho}^{[\g\d\e} {\rho}^{\h]\a\z} {f}_{\a\z}\,^{\b} = 0,
\end{equation}
the above condition is sufficient for the generalized fluxes of ExFT to be invariant under the generalized YB deformation \footnote{Although, at first sight this equation looks different from the equation obtained from the algebraic analysis \cite{Malek:2020hpo}, a careful inspection of generalized flux transformation under a tri-vector deformation shows that, for the case of group manifolds they are actually equivalent up to the unimodularity constraint~\eqref{uni}}. Given that the supergravity equations of motion can be written in terms of generalized fluxes and their derivatives, this proves that such tri-vector deformations always produce solutions of the usual 11d supergravity. Relaxing the unimodularity condition~\eqref{uni} one obtains a transformation of generalized fluxes parametrized by $J^{mn}$ and the corresponding generalized supergravity field equations, which are \emph{by construction} satisfied by the deformed backgrounds. 

After briefly describing our procedure, we will present the resulting generalized supergravity equations together with consistency conditions on the tensor $J^{mn}$. We will consider examples of non-trivial deformed backgrounds that solve these equations, but which are not solutions of the ordinary 11-dimensional supergravity. Details will be presented in an upcoming paper~\cite{Gubarev:2022}.

It is important to mention that at this stage the exact role played by these equations in the fundamental membrane dynamics is not clear. Our result certainly has two necessary features for \emph{the} theory of generalized 11d supergravity: i) It reproduces the equations of 10d generalized supergravity~\eqref{eq:gensugra} upon dimensional reduction; and ii) The tensor $J^{mn}$ appears in the field equations, carrying additional information about the background field configuration. However, until there are some further checks that show that these are also sufficient, it may be appropriate to think of what we have as \emph{a} tri-vector deformation of the supergravity equations.

\section{Generalized supergravity from double field theory}

Generalized supergravity can be obtained from the modified double field theory (DFT) construction of \cite{Sakatani:2016fvh}. Let us shortly review this construction in a form more suitable for generalization to 11-dimensions. DFT is a T-duality covariant representation of supergravity that by definition requires an extended space parametrized by the doubled set of $d+d$ coordinates $\XX^M = (x^m,\tx_m)$. 
In what follows we always assume that the fields do not depend on the dual coordinates $\tx_m$, thus the section constraint $\eta^{MN}\dt_M \otimes \dt_N = 0$ is satisfied. The extended space indices are raised and lowered using $\eta_{MN}$, which is the symmetric invariant tensor of $O(d,d)$. 

Bosonic sector of the theory is encoded in the generalized metric $\mH_{MN}$ parametrizing the coset space $O(d,d)/O(d)\times O(d)$, and the invariant dilaton $d = \Phi - \fr14 \log \det g_{mn}$. Local symmetries of the theory are the generalized diffeomorphisms, which act
\begin{equation}
    \label{eq:genlie}
    \begin{aligned}
         \mathbb{L}_\L V^M =&\  \L^N \dt_N V^M - V^N \dt_N \L^M + \h^{MN}\h_{KL}\dt_N \L^K V^L,\\
         \mathbb{L}_\L d  =&\ \L^M \dt_M d - \fr12 \dt_M \L^M
    \end{aligned}
\end{equation}
on a generalized vector $V^M$ and the dilaton $d$. These include standard diffeomorphisms, Kalb--Ramond gauge transformations and T-duality transformations.
For our purposes it is necessary to formulate the theory in terms of the generalized fluxes, or the anholonomicity coefficients $\mF_{AB}{}^C$
\begin{equation}
        [E_A,E_B]^M = \mF_{AB}{}^C E_C{}^M, \qquad
    \mathbb{L}_{E_A} d   = \fr12 \mF_A,
\end{equation}
where $E_A{}^M$ is the inverse of the generalized vielbein defined as usual as $\mH_{MN}=E_M{}^AE_N{}^B\mH_{AB}$ for a flat $O(d,d)$ metric $\mH_{AB}$. The fluxes satisfy generalized Bianchi identities
\begin{equation}
    \label{eq:BI_DFT}
    \begin{aligned}
        0&=\dt_{[A}\mF_{B C D]} - \fr34 \mF_{[AB}{}^E\mF_{CD]E},\\
        0&=2\dt_{[A}\mF_{B]}  + \dt^C \mF_{CAB}-\mF^C \mF_{CAB},\\
        0&=\dt^A \mF_A - \fr12\mF^A \mF_A + \fr1{12}\mF^{ABC}\mF_{ABC},
    \end{aligned}
\end{equation}
which may be understood as conditions of covariance under local transformations. 
The Lagrangian of the theory is then 
\begin{equation}\label{Lagrangian}
    \begin{aligned}
        e^{2d}\mL &=  \mH^{AB}\Big(2 \dt_A \mF_B - \mF_A \mF_B\Big) -2 \dt^A \mF_A + \mF^A\mF_A \\
        &+ \mF_{AB}{}^C \mF_{DE}{}^F \Big( \mH^{AD} \h^{BE} \h_{CF} + \mH^{AD}\mH^{BE}\mH_{CF}\Big) - \fr16 \mF_{ABC}\mF^{ABC},
    \end{aligned}
\end{equation}
where $\dt_A = E_A{}^M\dt_M$.

The generalized supergravity equations can be obtained by looking into the transformation of generalized fluxes under a bi-vector YB deformation, that is an $O (d,d)$ rotation $E_A{}^M \to O^M{}_N E_A{}^N$ with 
\begin{equation}
            O^M{}_N = \begin{bmatrix}
            \d^m{}_n && \beta^{m n} \\
            \\
            0 && \d_m{}^n
            \end{bmatrix}\in {O}(d,d).
\end{equation}
Given that the $r$-matrix defining the bi-vector  satisfies the classical YB equation \eqref{cyb}, the transformation of the fluxes reads
\begin{equation}\label{eq:fluxdefdft}
    \mF_{ABC}' = \mF_{ABC},\quad \mF_A' = \mF_A + X_A,
\end{equation}
where $X_A = E_A{}^M X_M =  E'{}_A{}^MX_M$ with $X_M = (0, I^m)$. In order to be able to interpret the deformed fluxes $\mF'_{ABC}$ and $\mF'_{A}$ as generalized fluxes of a new vielbein, we must ensure that the Bianchi identities~\eqref{eq:BI_DFT} hold both for the initial and the deformed fluxes. This results in the condition $\mathbb{L}_{X}{E'_{A}{}^M} = 0$,
i.e.\ it is enough to take $I^m = r^{\a\b}f_{\a\b}{}^\g k_\g^m$ to be a Killing vector of \emph{the deformed} background. It is important to mention, that generally speaking Bianchi identities produce both linear and quadratic constraints for $X^M$. The former gives the condition above, while the latter is $X^M X_M=0$ and is satisfied by the choice $X_M=(0,I^m)$. As will be seen later, such generalized Killing vector property of $X^M$ is a feature of bi-vector deformations only and does not hold for the 11-dimensional case.

To obtain the field equations for the NS-NS sector of generalized supergravity we start with DFT equations of motion in the flux formulation \cite{Geissbuhler:2013uka} that follow from \eqref{Lagrangian}
\begin{equation}
\begin{aligned}
    \label{eq:dfteqns} 
         0	&= (\mF_{C} - \dt_C)\mH^{D[A}\mF^{BC]}{}_D +2 \mH^{E [C}\mF_E{}^{DA}\mF_{CD}{}^{B]}  + \fr43  \mH^{C[A} \dt^{B]} {\cal F}_C   \\
         &- \fr23 \mH^{CE}\mH^{DF}\mH^{G[A}\mF_{EFG}\mF_{CD}{}^{B]}-\fr13 (\mF_{C} - \dt_C)\mH^{DC}\mH^{AE}\mH^{BF}\mF_{CEF},\\
         0 &= \mL\, ,     
\end{aligned}
\end{equation}
and according to \eqref{eq:fluxdefdft} substitute $\mF_A = \mF'_A - X_A$. The result will be a set of equations for the new vielbein $E'_M{}^A$, which are satisfied by construction given that $E_M{}^A$ satisfies \eqref{eq:dfteqns}. Explicit check shows that these are precisely the equations of generalized supergravity~\eqref{eq:gensugra}. 

The suggested approach is very close to the idea of deformed generalized Lie derivative considered in \cite{Ciceri:2016dmd,Cassani:2016ncu}, where the deformation is proportional to the Romans mass $m_R$ of the resulting massive Type IIA theory. The difference is that we deform not the Lie derivative but the fluxes, which results in conditions on the deformation rather than differential conditions on the fields. As
discussed in \cite{Ciceri:2016dmd}, the non-derivative terms in
the Lie derivative that determines the mass deformation can also
be obtained by a Scherk-Schwarz type ansatz. Likewise,
generalized supergravity equations in 10-dimensions can be
derived by imposing such an ansatz on certain fields either
within ExFT \cite{Baguet:2016prz} or DFT
\cite{Sakamoto:2017wor}. Hence, it is natural to expect that
the deformation induced from $J^{mn}$ to be derivable via
this mechanism. The relationship between these approaches will be investigated further in \cite{Gubarev:2022}.

\section{SL(5) case}

The DFT procedure described above can also be carried over to the case of tri-vector deformations, which is relevant to 11d backgrounds. Consider the SL(5) ExFT, which is a U-duality covariant formulation of supergravity in terms of the fields of the maximal $D=7$ supergravity \cite{Samtleben:2005bp}.  Its bosonic sector
\begin{equation}
    \begin{aligned}
        &\{g_{\m\n}, A_\m{}^{[MN]}, B_{\m\n\, M}, m_{MN}\}, && \m,\n=0,\dots,6,  \\
        & && M,N = 1,\dots,5,
    \end{aligned}
\end{equation}
contains the so-called external metric $g_{\m\n}$, generalized metric $m_{MN}\in \SL(5)/\SO(5)$ and 1- and 2-form fields $A_\m{}^{MN}, B_{\m\n \, M}$ transforming in the $\bf 10$ and $\bf \bar{5}$ of SL(5). All fields depend on $7+10$ coordinates $\{x^\m,X^{[MN]}\}$ and are subject to the section constraint
\begin{equation}
    \dt_{[MN}\otimes \dt_{KL]}=0.
\end{equation}
The theory is defined by the Lagrangian \cite{Hohm:2013pua,Musaev:2015ces}
\begin{equation}
    \begin{aligned}
     \mL=&\  \hat{\mc{R}} - \fr18 \mF_{\m\n}{}^{MN}\mF^{\m\n}{}_{ MN}+\frac{1}{48}h^{\m\n}\DD_{\m}m_{M N} \DD_\n m^{M N} \\
     &+ V(m,g)+{\fr{1}{3\cdot (16)^2}}m^{MN}\mF_{\m\n\r M}\mF^{\m\n\r}{}_N+\,\mL_{top}
    \end{aligned}
\end{equation}
where indices are raised and lowered by $m_{MN}$ and its inverse. The derivatives $\DD_\m = \dt_\m - \mathbb{L}_{A_\m}$ are covariant with respect to generalized diffeomorphisms defined as
\begin{equation}
    \begin{aligned}
   \mathbb{L}_{\Lambda} V^{M} =&\  \frac12 \Lambda^{KL} \partial_{KL}{V^{M}} - V^{L} \partial_{L K}{\Lambda^{M K}}
    + \frac14 V^{M} \partial_{K L}{\Lambda^{K L}}+ \lambda_V \dt_{KL}\L^{KL} V^M,
\end{aligned}
\end{equation}
when acting on a generalized vector $V^M$ of weight $\l_V$. To keep the setup as simple as possible we consider the following truncation of the theory \cite{Blair:2014zba,Bakhmatov:2020kul}:
\begin{equation}
    \begin{aligned}
        A_{\m}{}^{MN}&=0, \quad B_{\m\n\, M} =0,\\
        g_{\m\n}(x,X)&= e^{-2 \phi}h^{\fr15}\bar{g}_{\m\n}(x),\\
        m_{MN} & = e^{-\f}h^{\fr{1}{5}}M_{MN}.
    \end{aligned}
\end{equation}
Here $h= \det ||h_{mn}||$ denotes determinant of the $4\times 4$ block of the full 11-dimensional metric and the fields $\phi, h, M_{MN}$ are restricted to depend only on the extended coordinates $X^{[MN]}$. The $7\times 7$ block  $\bar{g}_{\m\n}$ of the full metric has only dependence on external coordinates. 

This allows to reformulate the theory in terms of only the metric $M_{MN}\in \SL(5)/\SO(5) \times \RR^+ $ or equivalently in terms of generalized fluxes $\mF_{AB,C}{}^D$ defined by
\begin{equation}
    \label{FluxF}
    \begin{aligned}
     \mathbb{L}_{E_{AB}}E^{M}{}_{C}&=\mF_{AB,C}{}^{D}E^{M}{}_{D}, \\ 
        \mF_{AB,C}{}^{D} & = \frac32 E_{N}{}^{D} \partial_{[A B} E^{N}{}_{C]} - E^{M}{}_{C} \partial_{M N} E^{N}{}_{[B} \delta^{D}{}_{A]} - \frac12 E^{M}{}_{[B|} \partial_{M N} E^{N}{}_{|A]} \delta^{D}{}_{C},
    \end{aligned}
\end{equation}
where $E_{A}{}^M$ is the inverse of the vielbein for the metric $M_{MN}$, $E_{AB}{}^{MN} = 4 E_{[A}{}^{M}E_{B]}{}^{N}$. We define $\dt_{AB} = E_{AB}{}^{MN}\dt_{MN}$ for clarity of notations.  The fluxes satisfy Bianchi identities
\begin{equation} 
\label{eq:bianchisl5}
\begin{aligned}
     0= & \  \frac32 \partial_{[A B}{\mF_{|D F| C]}{}^{E}} - \frac12 \partial_{D F}{\mF_{A B C}{}^{E}} +  \partial_{C G}{\mF_{D F [A}{}^{G}}\delta_{B]}{}^{E} - \frac14 \delta_{C}{}^{E} \partial_{B G}{\mF_{D F A}{}^{G}} + \frac14 \delta_{C}{}^{E} \partial_{A G}{\mF_{D F B}{}^{G}} \\
    &- \mF_{B G C}{}^{E} \mF_{D F A}{}^{G} + \mF_{A G C}{}^{E} \mF_{D F B}{}^{G}  + \mF_{A B G}{}^{E} \mF_{D F C}{}^{G} - \mF_{A B C}{}^{G} \mF_{D F G}{}^{E} .
\end{aligned}
\end{equation}
Finally, Lagrangian of the truncated theory in flux formulation reads
\begin{equation}
    \label{eq:lagrflux}
\begin{aligned}
m\ \mL'=& \ \bar{e} \, \mc{R}[\bar{g}_{(7)}] + Y_{A B} Y_{C D} m^{A C} m^{B D} - \fr12 Y_{A B} Y_{C D} m^{A B} m^{C D} \\
&+ 32  Z^{A B C} Z^{D E F} (m_{A D} m_{B E} m_{C F}+ m_{A C} m_{B D} m_{E F})- \frac{700}{3}\, {\q}_{A B} {\q}_{C D} {m}^{A C} {m}^{B D} ,
\end{aligned}
\end{equation}
where the fluxes $\q_{AB} \in \mathbf{10}, Y_{AB} \in \mathbf{15}, Z^{ABC} \in \mathbf{\overline{40}}$ are defined by
\begin{equation}
    \begin{aligned}
        \mF_{ABC}{}^D = 5 \theta_{[AB}\d_{C]}{}^D -2 \e_{ABCEF} Z^{EFD}  +  \d_{[A}{}^{D} Y_{B]C}.
    \end{aligned}
\end{equation} 
 Note, that the first two lines in \eqref{eq:lagrflux} similar in form to the scalar potential of $D=7$ maximal gauged supergravity, however here the fluxes are functions of the extended coordinates and hence the trombone contribution can be kept without violating the minimal action prescription.

Generalized Yang-Baxter deformation parametrized by a tri-vector  $\W^{mnk}= \fr1{3!}\r^{\a\b\g}k_{\a}{}^m k_{\b}{}^n k_{\g}{}^k$ is an SL(5) transformation~\cite{Bakhmatov:2020kul}
\begin{equation}
    \begin{aligned}
            O_M{}^N &= \begin{bmatrix}
            \d_m{}^n && 0 \\
            \\
            \fr1{3!}\e_{mpqr}\W^{pqr} && 1
            \end{bmatrix},
    \end{aligned}
\end{equation}
after which the flux components transform as
\begin{equation}\label{Fdeformation8}
\begin{aligned}
    \delta_{\rho} \mF_{ABC}{}^{D} = &\  {E}^{m}{}_{C} {E}^{n}{}_{A} {E}^{k}{}_{B} {E}_{l}{}^{D} {J}^{l p} {\epsilon}_{k m n p}  +\d_\rho^{(2)} \mF_{ABC}{}^{D} .
\end{aligned}
\end{equation}
Here we introduce the non-unimodularity parameter
\begin{equation}\label{Jbar}
J^{m n} = {\rho}^{\a\b\g} {f}_{\b\g}{}^\d\, {k}_{\a}{}^{m} {k}_{\d}{}^{n} = S^{m n} + I^{m n}, 
\end{equation}
that encodes terms in  the transformation linear in $\rho^{\a\b\g}$. $S^{mn}$ and $I^{mn}$ are the symmetric and antisymmetric parts of $J^{mn}$ respectively. The terms $\d_\rho^{(2)} \mF_{ABC}{}^{D}$ are quadratic in $\rho^{\a\b\g}$ and proportional to 
the generalized YB equation~\eqref{eq:sl5YB} that is equivalent to vanishing of the R-flux defined as $R^{p,mnkl} = 4\W^{pq[m}\nabla_{q}\W^{nkl]}$ \cite{Bakhmatov:2019dow}. Algebraically, this is the condition for  exceptional Drinfeld algebra to preserve its structure under a tri-vector deformation \cite{Sakatani:2019zrs}. Interestingly, it can be shown that setting $I^{mn}=0$ is sufficient for the generalized YB equation to be satisfied \cite{Bakhmatov:2019dow}. This is purely a matter of the low internal space dimension $d=4$ and the same is true for bi-vector YB deformations in $d=3$. Instead, we assume that the weaker quadratic condition is satisfied while $J^{mn}\neq 0$, which allows to arrive at a $d=11$ version of the generalized supergravity equations~\eqref{eq:gensugra}.

To do this, consider the generalized fluxes $    \mF'_{ABC}{}^D = \mF_{ABC}{}^D  + X_{ABC}{}^D,$
shifted precisely so as they transform under a non-unimodular generalized YB deformation, i.e.
\begin{equation}
    X_{mnk}{}^l = \epsilon_{mnkp}J^{lp}.
\end{equation}
One observes that $X_{MNK}{}^L$ cannot be interpreted as a generalized Killing vector like in the 10-dimensional case. Moreover, the non-unimodularity parameter $J^{mn}$ is essentially a $d=4$ tensor, hence, it breaks global $\GL(11)$ to $\GL(7)\times \GL(4)$. 

We now proceed with construction of the deformed theory. Following the analogy with the ten-dimensional case we consider quadratic and linear conditions following from the Bianchi identities~\eqref{eq:bianchisl5} on  $\mF'_{ABC}{}^D$ separately. For the former we have
\begin{equation}
    \label{eq:condse}
    \begin{aligned}
           L_{e_a}J^{kl} + J^{nl}\dt_n \phi\, e_a{}^k&=0, && &          J^{mn}\dt_n \phi&=0, \\
           \nabla_m\big(e^{-\f}I^{mn}\big) &= 0, && & J^{m[n}J^{kl]} &= 0,
    \end{aligned}
\end{equation}
which in particular implies the constraint $I^{[mn}I^{kl]}=0$ which is like a section condition. This gives a hint for the possibility to interpret the antisymmetric part $I^{mn}$ as a dual derivative of some field. These equations are straightforward generalization of the definition for $I^m$ \eqref{I} of the 10d case. The first equation above can be rewritten as $
    \nabla_{(m}J_{nk)}=0,$ which implies that the symmetric part $S^{mn} = J^{(mn)}$ is a Killing tensor of the deformed background. 

Additionally we have conditions that are linear in the gauge field $V^m = \frac{1}{3!} \ve^{mnkl}C_{nkl}$:
\begin{equation}
    \label{eq:condsV}
    \begin{aligned}
        \nabla_{[m}Z_{n]}  - \fr13 J^{kl}F_{mnkl}&=0,\\
        \nabla_k\Big(e^{-\phi}J^{k[l}V^{p]}\Big) &=0,\\
        \nabla_k(J^{(pl)}V^k) - \nabla_k(V^{(p}J^{l)k})&=0.
    \end{aligned}
\end{equation}
where
\begin{equation}
     Z_m = \dt_m \phi - \fr23 \ve_{mnkl}I^{nk}V^l,
\end{equation}
following a straightforward generalization of the ten-dimensional case. We also note, that the first condition can be nicely rewritten as 
\begin{equation}
    L_V I^{mn} +2 I^{p[m}V^{n]}\dt_p \phi=0,
\end{equation}
which together with the first two lines of \eqref{eq:condse} can be understood as action of $I^{mn}$ on the background via the Schouten--Nijenhuis bracket.

Equations of 11-dimensional supergravity generalized by adding a non-vanishing $J^{mn}$ then take the form
\begin{equation}
    \label{eq:gengensugra}
    \begin{aligned}
     0=&\ {\mR}_{m n}[h_{(4)}]  - 7\, \tilde{\nabla}_{(m}Z_{n)}  -\fr13 h_{mn}(\nabla V) + 8 (1+V^2)\Big(S_{mn}J^k{}_k - 2 J^{k}{}_{(m}J_{n) k}\Big)\\
     &+ 4 V_{m}V_{n}\Big(J^{kl}J_{kl} - 2 J^{kl}J_{lk}\Big) +4 V_kV_{l}\Big(4 J_{(m}{}^kJ_{n)}{}^l -J^k{}_{(m}J^l{}_{n)} - 2 S^{kl}S_{mn} \Big)  \\
     &+ 8 V_{k}V_{(m} \Big( 2J^l{}_{n)}J^k{}_l - 2  S_{n)}{}^k J^l{}_l  + J^{kl}J_{n)l}\Big),\\
    0=&\ \fr17 e^{2\f}\, \mc{R}[\bar{g}_{(7)}] + \frac{1}{6}\, (\nabla V)^2 +\tilde{\nabla}^m Z_m - 6 Z_m Z^m -2 J^{mn}J_{mn} + \fr43 J_{mn}J^{nm},\\
     0=&\ \tilde{\nabla}^m F_{mnkl} - 6 Z^m F_{mnkl}+ 6\Big(2 J^{pm}C_{m[nk}J_{l]p} - J^{pm}J_{p[n}C_{kl]m}\Big),
    \end{aligned}
\end{equation}
where $F_{mnkl} = 4 \dt_{[m}C_{nkl]}$ and 
\begin{equation}
    \begin{aligned}
        \tilde{\nabla}_m & = \nabla_m - \dt_m \phi \, .
    \end{aligned}
\end{equation}
The quantity $Z_m$ and the field strength $F_{mnkl}$ are required to satisfy 
\begin{equation}
    \nabla_{[m}Z_{n]} - \fr13I^{kl}F_{mnkl} =0.
\end{equation} 
When $J^{mn}=0$ these equations reproduce the truncated version of 11d supergravity equations given in \cite{Bakhmatov:2020kul}.

Terms in the equations above that are quadratic in $J^{mn}$ are the consequence of the fact that an analog of $X_m = I_m + Z_m$ of the 10-dimensional case cannot be defined here since $J^{mn}$ has two indices. Another reason for these to be expected is that after the dimensional reduction, various powers of $e^{\Phi}$ appear both in the Einstein and Maxwell equations $\phi$. It is easy to see that dimensional reduction of the generalized equations~\eqref{eq:gengensugra} reproduce the known result \eqref{eq:gensugra}. Indeed, suppose one of the Killing vectors $k_\a{}^m$ commutes with the remaining set forming a separate U(1) isometry to be identified with the M-theory circle. Then keeping non-zero only $I^{\bar{m}}= J^{4\bar{m}}\neq 0$ we are left with $X_{4\bar{m}\bar{n}}{}^4 \neq 0$, that can contribute only to $\mF_A$ simply by the index count. Finally, since the SL(5) theory after the reduction reproduces the O(3,3) DFT~\cite{Thompson:2011uw}, one just repeats the construction in the beginning of the letter. More details on the reduction will be given in an upcoming paper \cite{Gubarev:2022}.

\section{Examples}

The constraints \eqref{eq:condsV}
on $J^{mn}$ may look too restrictive, however, the theory is not void and contains non-trivial solutions. As an illustration let us consider two examples of non-unimodular generalized Yang-Baxter deformations of the $AdS_4 \times \SS^7$ solution. These are solutions to the equations \eqref{eq:gengensugra} by construction and the fields satisfy all the conditions above. We will use the Killing vectors of the $AdS_4$ space of radius $R$, which include three momentum generators $P_{a}$, dilatation $D$, angular momenta $M_{ab}$ and special conformal transformations $K_a$, 
where $a,b=0,1,2$. 

For our first example we consider the 3-vector
\begin{equation}
    \label{eq:def1}
    \W = \fr{2}{R^3}\,D\wedge (\r_a \e^{abc} P_b \wedge P_c).
\end{equation} 
Denoting $x_a = \h_{ab}x^b$, the deformed background becomes ($z$ is the radial AdS coordinate):
\begin{equation}
\begin{aligned}
ds^2 =&\ \fr{R^2}{4 z^2} K^{-\fr23} \left\{ dx_a dx^a -\fr1{z^2}\r_a dx^a dz  \left( 1+ \frac{\r_a x^a}{z^3} \right)dz^2 \right\}+ R^2 K^{\fr13} d\W_{(7)}^2,\\
F_{012z} =& -\fr38  \frac{R^3}{z^4} K^{-2}   \left( 1+ \frac{\r^2}{12 z^4} \right),\\
J^{ab} =& - \fr{4}{3R^3}\e^{abc}\r_{c},\\
K=&  \ 1+\dfrac{\r_a x^a}{z^3} - \dfrac{\r^2}{4z^4}.
\end{aligned}
\end{equation}
 For the $\rho$-tensor chosen as in \eqref{eq:def1} generalized YB equation \eqref{eq:sl5YB} is satisfied for arbitrary values of $\r_a$. In general the above does not solve equations of motion of the ordinary 11-dimensional supergravity, except for the special case $\r^2 = \r_a \r_b \h^{ab} = 0$ ~\cite{Bakhmatov:2020kul}. This simply means, that terms with and without $J^{mn}$ vanish separately, providing trivial solution to generalized equations in the sense of \cite{Wulff:2018aku}.

Our next example is a solution of the generalized theory only and corresponds to the deformation cubic in $x^a$:
\begin{equation}
\begin{aligned}
    \W & =    \frac{4}{R^3}\, \r_a \e^{abc}\, D\wedge M_{bd} \wedge M_c{}^d \\
     & =  \frac{4\r_a x^a}{R^3}\, \Big( x^b x_b \, \dt_0 \wedge \dt_1 \wedge \dt_2 - \fr z2 \, x_b \e^{bcd} \dt_c \wedge \dt_d\wedge \dt_z\Big). 
\end{aligned}
\end{equation}
The generalized YB equation \eqref{eq:sl5YB} constrains the $\r$-matrix as $\r^2 = \r_a \r_b \h^{ab} = 0$.
The background is then given by
\begin{equation}
\begin{aligned}
ds^2 =&\ \frac{R^2}{4 z^2}\, K^{-\fr23} \left\{ dx_a dx^a + \frac{1}{z^2}\, \r_a x^a x^b dx_b dz +\left( 1 - \frac{x_a x^a \r_b x^b}{z^3} \right) dz^2 \right\} + R^2 K^{\fr13} d\W_{(7)}^2,\\
F_{012z} =&\ -\fr38  \frac{R^3}{z^4} K^{-2} \left( 1 + \frac{1}{12} \frac{x_a x^a \r_b\r_c x^b x^c}{z^4} \right),\\
J^{ma} &= \fr{32}{R^3}\r_b\e^{abc}x_cx^m,\\
K =&\ 1 + \frac{x_a x^a}{z^3}\, \r_b x^b \, \left( 1 - \frac{\r_c x^c}{4z} \right),
\end{aligned}
\end{equation}
where $m=0,1,2,z$ and $a,b=0,1,2$.

\section{Conclusions}

We develop an effective method that allows to deform the equations of 11-dimensional supergravity by including an additional tensor $J^{mn}$ which is related to the non-unimodularity of tri-vector deformed 11d backgrounds, and is a generalization of the Killing vector $I^m$ \eqref{I}
of the 10d generalized supergravity. $J^{mn}$ \eqref{Jbar}
has similar properties to $I^m$, in particular, its symmetric part must be a Killing tensor $\nabla_{(m}J_{nk)} = 0$. Using this method we find a set of modified field equations of 11-dimensional supergravity \eqref{eq:gengensugra} which is an important step in connecting
10-dimensional generalized supergravity \cite{Arutyunov:2015mqj, Wulff:2016tju} to 11d. We also give two solutions that are obtained by the tri-vector generalized YB deformations. By construction, they are always solutions to the proposed equations \eqref{eq:gengensugra}.

Our algorithm is equally applicable to 10- and 11-dimensional supergravites, reproducing the known generalized supergravity equations \eqref{eq:gensugra} for the former. In 10d such possibility is allowed by the $\k$-symmetry constraints of the GS superstring and is associated with the breaking of conformal symmetry down to scale symmetry. The common lore is that a similar procedure cannot be done for the supermembrane due to the lack of conformal symmetry. However, our newly introduced tensor $J^{mn}$ breaks the global GL(11) diffeomorphism symmetry to GL(7)$\times$GL(4), allowing to avoid this obstruction. Hence, one expects that the fundamental supermembrane is $\k$-symmetric on solutions of \eqref{eq:gengensugra} at the cost of such global symmetry breaking. Showing this remains a task for future research.

\

{\bf Acknowledgments.} This work has been supported by Russian Science Foundation grant RSCF-20-72-10144. We would like to thank Gleb Arutyunov and David Berman for useful comments. NSD and EtM are grateful to Albert-Einstein-Institute, Potsdam, where this work was initiated. We thank IMBM, Istanbul for hospitality during the final phase of this work.


\begin{thebibliography}{10}

\bibitem{Arutyunov:2015mqj}
G.~Arutyunov, S.~Frolov, B.~Hoare, R.~Roiban, and A.~A. Tseytlin, ``{Scale
  invariance of the $\eta$-deformed $AdS_5\times S^5$ superstring, T-duality
  and modified type II equations},''
  \href{http://dx.doi.org/10.1016/j.nuclphysb.2015.12.012}{{\em Nucl. Phys.}
  {\bfseries B903} (2016) 262--303},
\href{http://arxiv.org/abs/1511.05795}{{\ttfamily arXiv:1511.05795 [hep-th]}}.

\bibitem{Wulff:2016tju}
L.~Wulff and A.~A. Tseytlin, ``{Kappa-symmetry of superstring sigma model and
  generalized 10d supergravity equations},''
  \href{http://dx.doi.org/10.1007/JHEP06(2016)174}{{\em JHEP} {\bfseries 06}
  (2016) 174},
\href{http://arxiv.org/abs/1605.04884}{{\ttfamily arXiv:1605.04884 [hep-th]}}.

\bibitem{Fernandez-Melgarejo:2018wpg}
J.~J. Fernández-Melgarejo, J.-I. Sakamoto, Y.~Sakatani, and K.~Yoshida,
  ``{Weyl invariance of string theories in generalized supergravity
  backgrounds},'' \href{http://dx.doi.org/10.1103/PhysRevLett.122.111602}{{\em
  Phys. Rev. Lett.} {\bfseries 122} no.~11, (2019) 111602},
\href{http://arxiv.org/abs/1811.10600}{{\ttfamily arXiv:1811.10600 [hep-th]}}.

\bibitem{Muck:2019pwj}
W.~M\"uck, ``{Generalized Supergravity Equations and Generalized
  Fradkin-Tseytlin Counterterm},''
  \href{http://dx.doi.org/10.1007/JHEP05(2019)063}{{\em JHEP} {\bfseries 05}
  (2019) 063}, \href{http://arxiv.org/abs/1904.06126}{{\ttfamily
  arXiv:1904.06126 [hep-th]}}.

\bibitem{Hoare:2015gda}
B.~Hoare and A.~A. Tseytlin, ``{On integrable deformations of superstring sigma
  models related to $AdS_n \times S^n$ supercosets},''
  \href{http://dx.doi.org/10.1016/j.nuclphysb.2015.06.001}{{\em Nucl. Phys.}
  {\bfseries B897} (2015) 448--478},
\href{http://arxiv.org/abs/1504.07213}{{\ttfamily arXiv:1504.07213 [hep-th]}}.

\bibitem{Hoare:2015wia}
B.~Hoare and A.~A. Tseytlin, ``{Type IIB supergravity solution for the T-dual
  of the $\eta$-deformed AdS$_{5} \times$ S$^{5}$ superstring},''
  \href{http://dx.doi.org/10.1007/JHEP10(2015)060}{{\em JHEP} {\bfseries 10}
  (2015) 060},
\href{http://arxiv.org/abs/1508.01150}{{\ttfamily arXiv:1508.01150 [hep-th]}}.

\bibitem{Araujo:2017jkb}
T.~Araujo, I.~Bakhmatov, E.~\'O~Colg\'ain, J.~Sakamoto, M.~M. Sheikh-Jabbari,
  and K.~Yoshida, ``{Yang-Baxter $\sigma$-models, conformal twists, and
  noncommutative Yang-Mills theory},''
  \href{http://dx.doi.org/10.1103/PhysRevD.95.105006}{{\em Phys. Rev.}
  {\bfseries D95} no.~10, (2017) 105006},
\href{http://arxiv.org/abs/1702.02861}{{\ttfamily arXiv:1702.02861 [hep-th]}}.

\bibitem{Bakhmatov:2017joy}
I.~Bakhmatov, O.~Kelekci, E.~\'O~Colg\'ain, and M.~M. Sheikh-Jabbari,
  ``{Classical Yang-Baxter Equation from Supergravity},''
  \href{http://dx.doi.org/10.1103/PhysRevD.98.021901}{{\em Phys. Rev.}
  {\bfseries D98} no.~2, (2018) 021901},
\href{http://arxiv.org/abs/1710.06784}{{\ttfamily arXiv:1710.06784 [hep-th]}}.

\bibitem{Borsato:2018idb}
R.~Borsato and L.~Wulff, ``{Non-abelian T-duality and Yang-Baxter deformations
  of Green-Schwarz strings},''
  \href{http://dx.doi.org/10.1007/JHEP08(2018)027}{{\em JHEP} {\bfseries 08}
  (2018) 027},
\href{http://arxiv.org/abs/1806.04083}{{\ttfamily arXiv:1806.04083 [hep-th]}}.

\bibitem{Bakhmatov:2018bvp}
I.~Bakhmatov and E.~T. Musaev, ``{Classical Yang-Baxter equation from
  $\beta$-supergravity},''
  \href{http://dx.doi.org/10.1007/JHEP01(2019)140}{{\em JHEP} {\bfseries 01}
  (2019) 140},
\href{http://arxiv.org/abs/1811.09056}{{\ttfamily arXiv:1811.09056 [hep-th]}}.

\bibitem{Klimcik:2008eq}
C.~Klim\v{c}{\'i}k, ``{On integrability of the Yang-Baxter sigma-model},''
  \href{http://dx.doi.org/10.1063/1.3116242}{{\em J. Math. Phys.} {\bfseries
  50} (2009) 043508},
\href{http://arxiv.org/abs/0802.3518}{{\ttfamily arXiv:0802.3518 [hep-th]}}.

\bibitem{Arutyunov:2013ega}
G.~Arutyunov, R.~Borsato, and S.~Frolov, ``{S-matrix for strings on
  $\eta$-deformed AdS5 x S5},''
  \href{http://dx.doi.org/10.1007/JHEP04(2014)002}{{\em JHEP} {\bfseries 04}
  (2014) 002}, \href{http://arxiv.org/abs/1312.3542}{{\ttfamily arXiv:1312.3542
  [hep-th]}}.

\bibitem{Arutyunov:2015qva}
G.~Arutyunov, R.~Borsato, and S.~Frolov, ``{Puzzles of $\eta$-deformed AdS$_5
  \times$ S$^5$},'' \href{http://dx.doi.org/10.1007/JHEP12(2015)049}{{\em JHEP}
  {\bfseries 12} (2015) 049},
\href{http://arxiv.org/abs/1507.04239}{{\ttfamily arXiv:1507.04239 [hep-th]}}.

\bibitem{Bergshoeff:1987cm}
E.~Bergshoeff, E.~Sezgin, and P.~K. Townsend, ``{Supermembranes and
  Eleven-Dimensional Supergravity},''
  \href{http://dx.doi.org/10.1016/0370-2693(87)91272-X}{{\em Phys. Lett.}
  {\bfseries B189} (1987) 75--78}.
[,69(1987)].

\bibitem{Howe:1997he}
P.~S. Howe, ``{Weyl superspace},''
  \href{http://dx.doi.org/10.1016/S0370-2693(97)01261-6}{{\em Phys. Lett.}
  {\bfseries B415} (1997) 149--155},
\href{http://arxiv.org/abs/hep-th/9707184}{{\ttfamily arXiv:hep-th/9707184
  [hep-th]}}.

\bibitem{Berman:2020tqn}
D.~S. Berman and C.~D.~A. Blair, ``{The Geometry, Branes and Applications of
  Exceptional Field Theory},''
  \href{http://dx.doi.org/10.1142/S0217751X20300148}{{\em Int. J. Mod. Phys. A}
  {\bfseries 35} no.~30, (2020) 2030014},
  \href{http://arxiv.org/abs/2006.09777}{{\ttfamily arXiv:2006.09777
  [hep-th]}}.

\bibitem{Baguet:2015xha}
A.~Baguet, O.~Hohm, and H.~Samtleben, ``{E$_{6(6)}$ Exceptional Field Theory:
  Review and Embedding of Type IIB},'' {\em PoS} {\bfseries CORFU2014} (2015)
  133,
\href{http://arxiv.org/abs/1506.01065}{{\ttfamily arXiv:1506.01065 [hep-th]}}.

\bibitem{Gubarev:2020ydf}
K.~Gubarev and E.~T. Musaev, ``{Polyvector deformations in eleven-dimensional
  supergravity},'' \href{http://dx.doi.org/10.1103/PhysRevD.103.066021}{{\em
  Phys. Rev. D} {\bfseries 103} no.~6, (2021) 066021},
  \href{http://arxiv.org/abs/2011.11424}{{\ttfamily arXiv:2011.11424
  [hep-th]}}.

\bibitem{Bakhmatov:2019dow}
I.~Bakhmatov, N.~S. Deger, E.~T. Musaev, E.~\'O~Colg\'ain, and M.~M.
  Sheikh-Jabbari, ``{Tri-vector deformations in $d=11$ supergravity},''
  \href{http://dx.doi.org/10.1007/JHEP08(2019)126}{{\em JHEP} {\bfseries 08}
  (2019) 126},
\href{http://arxiv.org/abs/1906.09052}{{\ttfamily arXiv:1906.09052 [hep-th]}}.

\bibitem{Malek:2020hpo}
E.~Malek, Y.~Sakatani, and D.~C. Thompson, ``{E$_{6(6)}$ exceptional
  Drinfel\textquoteright{}d algebras},''
  \href{http://dx.doi.org/10.1007/JHEP01(2021)020}{{\em JHEP} {\bfseries 01}
  (2021) 020}, \href{http://arxiv.org/abs/2007.08510}{{\ttfamily
  arXiv:2007.08510 [hep-th]}}.

\bibitem{Gubarev:2022}
I.~Bakhmatov, A.~Catal-Ozer, N.~S. Deger, K.~Gubarev, and E.~T. Musaev, ``{to
  appear},'' {\em to appear} (2022) .

\bibitem{Sakatani:2016fvh}
Y.~Sakatani, S.~Uehara, and K.~Yoshida, ``{Generalized gravity from modified
  DFT},'' \href{http://dx.doi.org/10.1007/JHEP04(2017)123}{{\em JHEP}
  {\bfseries 04} (2017) 123},
\href{http://arxiv.org/abs/1611.05856}{{\ttfamily arXiv:1611.05856 [hep-th]}}.

\bibitem{Geissbuhler:2013uka}
D.~Geissbuhler, D.~Marques, C.~Nunez, and V.~Penas, ``{Exploring Double Field
  Theory},'' \href{http://dx.doi.org/10.1007/JHEP06(2013)101}{{\em JHEP}
  {\bfseries 06} (2013) 101}, \href{http://arxiv.org/abs/1304.1472}{{\ttfamily
  arXiv:1304.1472 [hep-th]}}.

\bibitem{Ciceri:2016dmd}
F.~Ciceri, A.~Guarino, and G.~Inverso, ``{The exceptional story of massive IIA
  supergravity},'' \href{http://dx.doi.org/10.1007/JHEP08(2016)154}{{\em JHEP}
  {\bfseries 08} (2016) 154}, \href{http://arxiv.org/abs/1604.08602}{{\ttfamily
  arXiv:1604.08602 [hep-th]}}.

\bibitem{Cassani:2016ncu}
D.~Cassani, O.~de~Felice, M.~Petrini, C.~Strickland-Constable, and D.~Waldram,
  ``{Exceptional generalised geometry for massive IIA and consistent
  reductions},'' \href{http://dx.doi.org/10.1007/JHEP08(2016)074}{{\em JHEP}
  {\bfseries 08} (2016) 074}, \href{http://arxiv.org/abs/1605.00563}{{\ttfamily
  arXiv:1605.00563 [hep-th]}}.

\bibitem{Baguet:2016prz}
A.~Baguet, M.~Magro, and H.~Samtleben, ``{Generalized IIB supergravity from
  exceptional field theory},''
  \href{http://dx.doi.org/10.1007/JHEP03(2017)100}{{\em JHEP} {\bfseries 03}
  (2017) 100},
\href{http://arxiv.org/abs/1612.07210}{{\ttfamily arXiv:1612.07210 [hep-th]}}.

\bibitem{Sakamoto:2017wor}
J.-i. Sakamoto, Y.~Sakatani, and K.~Yoshida, ``{Weyl invariance for generalized
  supergravity backgrounds from the doubled formalism},''
  \href{http://dx.doi.org/10.1093/ptep/ptx067}{{\em PTEP} {\bfseries 2017}
  no.~5, (2017) 053B07},
\href{http://arxiv.org/abs/1703.09213}{{\ttfamily arXiv:1703.09213 [hep-th]}}.

\bibitem{Samtleben:2005bp}
H.~Samtleben and M.~Weidner, ``{The Maximal D=7 supergravities},''
  \href{http://dx.doi.org/10.1016/j.nuclphysb.2005.07.028}{{\em Nucl.Phys.}
  {\bfseries B725} (2005) 383--419},
\href{http://arxiv.org/abs/hep-th/0506237}{{\ttfamily arXiv:hep-th/0506237
  [hep-th]}}.

\bibitem{Hohm:2013pua}
O.~Hohm and H.~Samtleben, ``{Exceptional Form of D=11 Supergravity},''
  \href{http://dx.doi.org/10.1103/PhysRevLett.111.231601}{{\em Phys.Rev.Lett.}
  {\bfseries 111} (2013) 231601},
\href{http://arxiv.org/abs/1308.1673}{{\ttfamily arXiv:1308.1673 [hep-th]}}.

\bibitem{Musaev:2015ces}
E.~T. Musaev, ``{Exceptional field theory: $SL(5)$},''
  \href{http://dx.doi.org/10.1007/JHEP02(2016)012}{{\em JHEP} {\bfseries 02}
  (2016) 012},
\href{http://arxiv.org/abs/1512.02163}{{\ttfamily arXiv:1512.02163 [hep-th]}}.

\bibitem{Blair:2014zba}
C.~D.~A. Blair and E.~Malek, ``{Geometry and fluxes of SL(5) exceptional field
  theory},'' \href{http://dx.doi.org/10.1007/JHEP03(2015)144}{{\em JHEP}
  {\bfseries 03} (2015) 144},
\href{http://arxiv.org/abs/1412.0635}{{\ttfamily arXiv:1412.0635 [hep-th]}}.

\bibitem{Bakhmatov:2020kul}
I.~Bakhmatov, K.~Gubarev, and E.~T. Musaev, ``{Non-abelian tri-vector
  deformations in $d=11$ supergravity},''
  \href{http://dx.doi.org/10.1007/JHEP05(2020)113}{{\em JHEP} {\bfseries 05}
  (2020) 113}, \href{http://arxiv.org/abs/2002.01915}{{\ttfamily
  arXiv:2002.01915 [hep-th]}}.

\bibitem{Sakatani:2019zrs}
Y.~Sakatani, ``{$U$-duality extension of Drinfel'd double},''
  \href{http://dx.doi.org/10.1093/ptep/ptz172}{{\em PTEP} {\bfseries 2020}
  no.~2, (2020) 023B08}, \href{http://arxiv.org/abs/1911.06320}{{\ttfamily
  arXiv:1911.06320 [hep-th]}}.

\bibitem{Thompson:2011uw}
D.~C. Thompson, ``{Duality Invariance: From M-theory to Double Field Theory},''
  \href{http://dx.doi.org/10.1007/JHEP08(2011)125}{{\em JHEP} {\bfseries 1108}
  (2011) 125},
\href{http://arxiv.org/abs/1106.4036}{{\ttfamily arXiv:1106.4036 [hep-th]}}.

\bibitem{Wulff:2018aku}
L.~Wulff, ``{Trivial solutions of generalized supergravity vs non-abelian
  T-duality anomaly},''
  \href{http://dx.doi.org/10.1016/j.physletb.2018.04.025}{{\em Phys. Lett. B}
  {\bfseries 781} (2018) 417--422},
  \href{http://arxiv.org/abs/1803.07391}{{\ttfamily arXiv:1803.07391
  [hep-th]}}.

\end{thebibliography}

\providecommand{\href}[2]{#2}\begingroup\raggedright\endgroup

\end{document}